\documentstyle[11pt,a4,amssymb]{article}

\newcommand{\be}{\begin{equation}}
 \newcommand{\ee}{\end{equation}} \newcommand{\bea}{\begin{eqnarray}}
 \newcommand{\eea}{\end{eqnarray}}

  \catcode`\@=11 \@addtoreset{equation}{section}
  \catcode`\@=11

\begin{document}

\begin{titlepage}

\begin{flushright} {\tt FTUV/95-31\\ IFIC/95-33\\ hep-th/9512187}
 \end{flushright}

\vfill

\begin{center}

{\bf{\Large Solvable Models for radiating
 Black Holes \\and Area-preserving
 Diffeomorphisms\footnote{Work partially supported by the {\it
Comisi\'on Interministerial de Ciencia y Tecnolog\'{\i}a}\/ and {\it DGICYT}.}}
}

\bigskip \bigskip \bigskip J.Cruz\footnote{\sc cruz@lie.ific.uv.es}and J.
Navarro-Salas\footnote{\sc jnavarro@lie.ific.uv.es}

\bigskip

\begin{center} Departamento de F\'{\i}sica Te\'orica and\\ IFIC, Centro Mixto
	Universidad de Valencia-CSIC.\\ Facultad de F\'{\i}sica, Universidad de
	Valencia,\\ Burjassot-46100, Valencia, Spain.\\

\end{center} \bigskip \today \bigskip

\end{center}


\begin{center} {\bf Abstract} \end{center}
Solvable theories of 2D dilaton
gravity can be obtained from a Liouville theory by suitable field
redefinitions.
In this paper we propose a new framework to generate 2D dilaton gravity models
which can also be exactly solved in the semiclassical approximation.  Our
approach is based on the recently introduced scheme to quantize massless scalar
fields coupled to 2D gravity maintaining
invariance under area-preserving diffeomorphisms and
Weyl transformations.  Starting from the CGHS model with the new effective
action we
reestablish the full diffeomorphism invariance by means of an adequate family
of field redefinitions.  The original theory is therefore mapped into a large
family of solvable models .We focus our analysis on
 the one-parameter class of
models interpolating between the Russo-Susskind-Thorlacius model and the
Bose-Parker-Peleg model.  Finally we shall briefly indicate how can we extend
our approach to spherically symmetric Einstein gravity coupled to 2D conformal
matter.

\vfill \end{titlepage} \newpage

\section{Introduction} One of the most interesting features of 2D dilaton black
 holes is the possibility of constructing solvable models for the evaporation
 process including the back-reaction effects. The solvability of the models can
 be traced back to the fact that, in conformal gauge, a
 set of field redefinitions of the
 metric-dilaton fields convert the gravity models into a Liouville theory
 \cite{Bilal.Callan}. A simple model of this type was provided by
 Russo, Susskind and Thorlacius \cite{RST}. The addition of the counterterm
 \be
 -{N\over96\pi} \int d^2 x\sqrt{-g}2\phi R \>,\label{i}
  \ee
   to the standard
 one-loop effective action of the Callan-Giddings-Harvey-
 Strominger (CGHS) model \cite{CGHS} $S_0+S_P$ , where
  \be
  S_0={1\over2\pi}\int d^2
 x\sqrt{-g}\left[e^{-2\phi}\left(R+4\left(\nabla\phi\right
 )^2+4\lambda^2\right)-{1\over2}\sum_{i=1}^N\left(\nabla f_i\right)^2\right]
  \>,\label{ii}
  \ee
  is the classical CGHS action and $S_P$ is the Polyakov action
   \be
 S_P=-{N\over96\pi}\int d^2x\sqrt{-g}R\square^{-1}R\>,\label{iii} \ee restores
 the solvability of the classical theory. The field redefinitions
 $(k={N\over12})$:
 \be
 \Omega={\sqrt{k}\over2}\phi+{e^{-2\phi}\over\sqrt{k}}\>,\label{iv} \ee \be
 \chi=\sqrt{k}\rho-{\sqrt{k}\over2}\phi+{e^{-2\phi}\over\sqrt{k}} \>,
 \label{v}
 \ee
 convert the RST model \cite{RST} into a Liouville theory
 \be
 {1\over\pi}
 \int d^2x  \left[
 -\partial_+\chi\partial_-\chi
 +\partial_+\Omega\partial_-\Omega
 +\lambda^2 e^{{2\over\sqrt{k}}
 \left(
 \chi-
 \Omega \right)}	
 + {1\over2} \sum_{i=1}^N \partial_+ f_i \partial_- f_i \right]
 \>.
 \label{vi}
 \ee
 Under general assumptions: i) the
 absence of unphysical fluxes at infinity in the vacuum, ii)
  the existence of a
 vacuum solution with $R=0$, the most general semiclassical
 theory, containing the standard Polyakov term, that can be cast into the
 form (\ref{vi}) is given by the one-parameter class of
 models \cite{Fabbri-Russo}
 \bea && S={1\over2\pi}\int d^2x\sqrt{-g}\left[
 e^{-{2\over n}\phi}\left(R+{4\over n}\left
 (\nabla\phi\right)^2\right)+4\lambda^2e^{-2\phi}-{1\over2}\sum_{i=1}^N
 \left(\nabla
 f_i\right)^2\right.\nonumber\\
 &&\left.+k\left({1-2n\over2n}\phi R+{\left(n-1\right)\over n}
 \left(\nabla\phi
 \right)^2-{1\over4}R\square^{-1}R\right)\right]\>,
 \label{vii}
 \eea
 where the
 different models are labelled by the parameter n.

   The above family of dilaton-gravity models is however rather
    restrictive. The interesting case of
  Einstein theory -restricted to spherical field configurations-
   coupled to null matter
  $\left(^{\left(4\right)}ds^2=g_{\mu\nu}dx^{\mu}dx^{\nu}
  +{e^{-2\phi}\over\lambda^2}d\Omega^2\ \mu,\nu=1,2\right)$
\be
 {1\over2\pi}\int d^2x\sqrt{-g}\left[e^{-2\phi}\left(R+2
 \left(\nabla\phi
\right)^2+2\lambda^2e^{2\phi}\right)-{1\over2}\sum_{i=1}^N\left(
\nabla f_i\right)^2\right]\>,\label{viii}
 \ee
  cannot be reached by this framework.

In this paper we want to put forward an alternative approach to construct
exactly solvable models by extending the results of Ref \cite{Navarro}.
In section 2 we introduce the basic ingredients of the Weyl-invariant
quantization scheme for two-dimensional massless fields. In section 3
we describe a procedure of field redefinitions to generate exactly
solvable semiclassical models for the CGHS theory. Section 4 is devoted
to study the one-parameter class of solvable models which continuously
interpolates between the RST model and the Bose-Parker-Peleg (BPP) model
\cite{BPP}. In section 5 we generalize the structure of the field
redefinitions of the CGHS theory to a general class of dilaton-gravity
models and define the analogue of the RST and BPP models
for spherically symmetric gravity.
\section {Area-preserving diffeomorphisms and the Virasoro anomaly}
 The Polyakov
effective action $S_P$ in (\ref{ii}) reflects the well known fact that the
expectation value of the stress tensor operator of
N massless scalar fields coupled to gravity in two dimensions verify the
equations:
\be
 \nabla_{\mu}<T^{f\mu\nu}>=0\>,\label{ix}
\ee
 \be
<T^{f\mu}_{\ \mu} >={N\over24}R\>.\label{x}
 \ee
 The covariant conservation
equation (\ref{ix}) is a consequence of the covariance in the choice of
 the functional measure for the
conformal fields:
\be
\parallel\delta f_i\parallel^2=\int d^2x\sqrt{-g}\delta f_i\delta f_i
\>.\label{xi}
\ee
 The anomaly relation (\ref{x}) reflects
  the breaking of Weyl symmetry in the quantization procedure.
   One can alternatively propose a different way of defining the functional
measure that preserve Weyl invariance \cite{Karakhaniyan,Jackiw}
(see also \cite{Amelino} ).
\be
\parallel\delta f_i\parallel^2=\int d^2x\delta f_i\delta f_i\>.\label{measure}
\ee
Diffeomorphism invariance is therefore
lost although invariance under area-preserving diffeomorphisms
 is maintained (see \cite{H.S.La} for a
general study). This leads to a new effective action $S_W$
 for conformal matter
 \be
S_W\left(g^{\mu\nu}\right)=S_P\left(\sqrt{-g}g^{\mu\nu}\right)\>,
\label{xii}
 \ee
 and the new effective stress tensor
 \be
T_{\mu\nu}^W=-{2\pi\over\sqrt{-g}}{\delta S_W\over\delta g^{\mu\nu}}
\>,\label{xiii}
\ee
 satisfies the anomaly relations
  \be
T^{W\mu}_{\ \mu}=0\>,\label{xiv}
 \ee
 \be
 \nabla^{\mu}T^W_{\mu\nu}=-{N\over48}
 {1\over\sqrt{-g}}\partial_{\nu}R\left(\sqrt{-g}
g^{\mu\nu}\right)\>,\label{xv}
 \ee
  In conformal
gauge, $ds^2=-e^{2\rho}dx^+dx^-$,
the above expressions take a simple form
\be
T^W_{+-}=0\>,\label{xvi}
 \ee
  \be
\nabla^+T_{++}^W=\nabla^-T^W_{--}=0\>.\label{xvii}
 \ee
 $T^W_{\mu\nu}$ transforms
covariantly under area-preserving diffeomorphisms, but under general conformal
coordinate transformations $x^{\pm}\longrightarrow y^{\pm}\left(
x^{\pm}\right)\ T_{\pm\pm}^W$ transforms
according to the Virasoro anomaly:
\be
T^W_{x^{\pm} x^{\pm}} = \left( {dy^{\pm} \over dx^{\pm}} \right)^2
T^W_{y^{\pm} y^{\pm}} - {N\over24} \left\{ y^{\pm}, x^{\pm}\right\} \>,
 \label{xviii}\ee
where $\left\{y,x\right\}$ is the schwartzian derivative.
 \section{Field redefinitions and
solvable models for the CGHS theory}
 The relationship between the non-covariant scheme defined by
 the "semiclassical" action
  \be
S_0+S_W\>,\label{xix}
\ee
and the
standard (covariant) approach was investigated
in Ref \cite{Navarro}. By relating in conformal gauge
the solutions of the non-covariant theory
 with
the classical solutions in vacuum it was found
 a set of field redefinitions
  which
convert the theory (\ref{xix}) into the RST model
 . The field
redefinitions are:
  \be
   \rho-\phi=\hat\rho-\hat\phi\>,\label{xx}
    \ee
 \be
e^{-2\phi}=e^{-2\hat\phi}+{N\over24}\hat\phi
+{N\over12}M\left(\hat\rho-\hat\phi
\right)\>,\label{xxi}
 \ee
  where
\begin{eqnarray*}
M\left(\hat\rho-\hat\phi\right)&=&\hat\rho-\hat\phi + \int^{y^+}
e^{2\left(\hat\rho-\hat\phi\right)}
 \int e^{-2\left(\hat\rho-\hat\phi\right)}
  \left[ \partial_{y^+}
\left(\hat\rho-\hat\phi\right) \right]^2 \nonumber \\ && + \int^{y^-}
e^{2\left(\hat\rho-\hat\phi\right)}
 \int e^{-2\left(\hat\rho-\hat\phi\right)} \left[ \partial_{y^-}
\left(\hat\rho-\hat\phi\right) \right]^2
\end{eqnarray*}
$\left(\phi,\rho\right)$ and $\left(\hat\phi,\hat\rho\right)$
 refer to the non-covariant and covariant theories
respectively.

 Now we shall further explore the structure of the above field
redefinitions.
The anomalous transformation properties of the fields
$\left(\phi,\rho\right)$ can be read immediately
from the form of the classical solution
\bea
e^{-2\phi}&=&-\int^{y^+}e^{2\left(\rho-\phi\right)}\int
e^{-2\left(\rho-\phi\right)} \left(T_{++}^{cl}+T_{++}^W\right)\nonumber\\
&&-\int^{y^-}e^{2\left(\rho-\phi\right)}
\int e^{-2\left(\rho-\phi\right)} \left(
T_{--}^{cl}+T_{--}^W\right)\nonumber\\
&&-\lambda^2e^{-2\left(
\rho-\phi\right)}\int^{y^+}e^{2\left(\rho-\phi\right)}
\int^{y^-}e^{2\left(\rho-\phi\right)}\>.\label{xxii}
\eea
The Virasoro anomaly of $T_{\pm\pm}^W$ makes that, under the
conformal transformations $y^{\pm}
\longrightarrow z^{\pm}\left(y^{\pm}\right),\ e^{-2\phi}$ behaves as
\be
e^{-2\phi}\longrightarrow e^{-2\phi}-{N\over24}\int e^{2\left(\rho-\phi\right)}
\int e^{-2\left(\rho-\phi\right)}\left({dz^+\over dy^+}\right)^2
\left[{d^2\over dz^{+2}}\log
{dy^+\over dz^+}-{1\over2}\left({d\over dz^+}\log
{dy^+\over dz^+}\right)^2\right]\>.\label{xxiii}
\ee
This anomalous transformation is just cancelled by the
 transformation law of the term ${N\over12}M\left(\hat\rho
-\hat\phi\right)$ when $\left(\hat\phi,\hat\rho\right)$ transform covariantly.
This suggests the following generalization of
 the redefinitions
(\ref{xx}),(\ref{xxi}):
\be
\rho-\phi=\hat\rho-\hat\phi\>,\label{xxiv}
\ee
\be
e^{-2\phi}=e^{-2\hat\phi}+{N\over12}F\left(\hat\phi\right)+{N\over12}
M\left(\hat\rho-\hat\phi\right)\>,\label{xxv}
\ee
where F is an arbitrary function. It is clear that the new redefinitions
still transform the non-covariant theory (\ref{xix}) into a covariant one.
We must point out that the term $M\left(\hat\rho-\hat\phi\right)$ is a
purely gauge-dependent term since its argument is a conserved current.
The equations of motion derived from the non-covariant action (\ref{xix})
are
\be
\partial_+\partial_-e^{-2\phi}+\lambda^2e^{2\left(\rho-\phi\right)}=0
\>,\label{xxvi}
\ee
\be
\partial_+\partial_-\left(\rho-\phi\right)=0\>,\label{xxvii}
\ee
\be
\partial^2_{\pm}e^{-2\phi}-2\partial_{\pm}\left(\rho-\phi\right)
\partial_{\pm}e^{-2\phi}+T^{f}_{\pm\pm}+T^{W}_{\pm\pm}=0\>.\label{xxviii}
\ee
The field redefinitions (\ref{xxiv}),(\ref{xxv}) convert
the above non-covariant equations
into the covariant ones
\be
\partial_+\partial_-\left(e^{-2\hat\phi}+{N\over12}F\left(\hat\phi\right)
\right)+\lambda^2e^{2\left(\hat\rho-\hat\phi\right)}=0\>,\label{xxix}
\ee
\be
\partial_+\partial_-\left(\hat\rho-\hat\phi\right)=0\>,\label{xxx}
\ee
\bea
&&\partial^2_{\pm}\left(e^{-2\hat\phi}+{N\over12}F\left(\hat\phi\right)\right)
-2\partial_{\pm}\left(\hat\rho-\hat\phi\right)\partial_{\pm}\left(
e^{-2\hat\phi}+{N\over12}F\left(\hat\phi\right)\right)\nonumber\\
&&-{N\over12}\left[\left(\partial_{\pm}\left(\hat\rho-\hat\phi\right)
\right)^2-\partial^2_{\pm}\left(\hat\rho-\hat\phi\right)\right]+T^f
_{\pm\pm}+T^W_{\pm\pm}=0\>.\label{xxxi}
\eea
It is not difficult to check that this equations can be obtained from the
following action
\be
S_0+S_P+{N\over24\pi}\int d^2x\sqrt{-g}\left[\left(\nabla\phi\right)^2
-2g^{\mu\nu}\nabla_{\mu}F\left(\phi\right)\nabla_{\nu}\phi+F\left(\phi\right)R
-\phi R\right]\>,\label{xxxii}
\ee
which therefore defines a large family of 2D dilaton-gravity
models parametrized by the arbitrary function F. Since the non-covariant theory
is solvable the models (\ref{xxxii}) are automatically solvable.
In the so-called "Kruskal" gauge $\hat\rho=\hat\phi$ the gauge dependent term
$M\left(\hat\rho-\hat\phi\right)$ vanishes and then the
solutions to the semiclassical equations
 become
\be
e^{-2\hat\phi}+{N\over12}F\left(\hat\phi\right)=e^{-2\phi}\>,\label{xxxiii}
\ee
where $\phi$ is the solution (\ref{xxii}) of the non-covariant theory
\be
e^{-2\phi}=e^{-2\phi_{cl}}-{N\over48}\log\left(-\lambda^2x^+x^-\right)
\>,\label{xxxiv}
\ee
$\phi_{cl}$ is the corresponding classical solution and $x^{\pm}$ are the
coordinates defined by the gauge $\hat\phi=\hat\rho$.
The models (\ref{vii}) can be recovered if one choose the function F as
follows:
\be
F\left(\hat\phi\right)={12\over N}\left(e^{-{2\over n}\hat \phi}-e^{-2\hat\phi}
\right)+{\hat\phi\over 2n}\>.\label{xxxv}
\ee
Observe that the above choice modifies the classical action.
Another natural choice for the function F is
\be
F\left(\hat\phi\right)=a\hat\phi\>,\label{xxxvi}
\ee
where $a$ is an arbitrary real parameter. In fact one can show that the above
 models are the only ones that can obtained from the procedure of Ref
 \cite{Navarro} by admitting a general form for the boundary conditions
 $T^W_{x^{\pm}x^{\pm}}$. The requirement
 of being able to rewrite the non-covariant semiclassical
 vacuum solutions in terms of the classical ones implies
  that $T^W_{x^{\pm}x^{\pm}}=-{Na\over24}{1\over{\left(x^{\pm}\right)^2}}$.
 Hence one obtain the redefinitions (\ref{xxiv}),(\ref{xxv})
  with $F\left(\hat\phi
 \right)=a\hat\phi$. The semiclassical action (\ref{xxxii}) then becomes
 \be
 S_0+S_P+{N\over24\pi}\int d^2 x\sqrt{-g}\left[\left(1-2a\right)
 \left(\nabla\phi\right)^2+\left(a-1\right)\phi R\right]\>.\label{xxxvii}
 \ee
 This one-parameter class of models continuously interpolates between the
 RST model $\left(a={1\over2}\right)$ and the recently introduced model of
 Bose-Parker-Peleg \cite{BPP} (BPP-model) $\left(a=0\right)$ which
  describes an evaporating
 black hole having a remnant as the semiclassical end-state geometry.
 The later model has recently emerged in the semiclassical limit of
 the non-perturbative approach of Ref. \cite{Mikovic}.

 To finish this section let us briefly mention that the general semiclassical
 action (\ref{xxxii}) can also be transformed to the Liouville theory
 (\ref{vi}).
 The field redefinitions that
  reduce (\ref{xxxii}) to (\ref{vi}) in conformal gauge are
 \be
 \Omega=\sqrt{k}F\left(\phi\right)+{e^{-2\phi}\over\sqrt{k}}\>,\label{xxxviii}
 \ee
 \be
 \chi=\sqrt{k}\rho+\sqrt{k}\left(F\left(\phi\right)-\phi\right)+
 {e^{-2\phi}\over\sqrt{k}}\>,\label{xxxix}
 \ee
 \section{The RST-BPP models}
In this section we shall briefly consider the gravitational collapse described
by the one-parameter class of models (\ref{xxxvii})
thus interpolating the analysis of
\cite{RST} and \cite{BPP}. First we shall study the
solutions with no radiation at infinity $<T^f_{\pm\pm}\left(\sigma^{\pm}
\right)>=0$. Taking into account (\ref{xxxiii}),(\ref{xxxiv})
   with $F\left(\phi\right)=a\phi$ we find
that the asymptotically Minkowski vacuum solution, in Kruskal gauge, is
\be
e^{-2\phi}+{Na\over12}\phi=-\lambda^2x^+x^--{N\over48}\log\left(-\lambda^2
x^+x^-\right)+C\>,\label{xl}
\ee
where C is an integration constant. In the coordinates $\sigma^{\pm}$, we
have\be
e^{-2\phi}+{Na\over12}\phi=e^{2\lambda\sigma}-{N\over48}2\lambda\sigma+C
\>.\label{xli}\ee
For $\sigma\longrightarrow\infty$ the expression (\ref{xli})
and the analogue
one for the metric
\be
e^{-2\rho}+{Na\over12}e^{-2\lambda\sigma}\left(\rho-\lambda\sigma\right)
=1-e^{-2\lambda\sigma}\left({N\over48}2\lambda\sigma+C\right)\>,\label{xlii}
\ee
approach to the linear dilaton vacuum.
Let us now consider the solution describing a collapsing shell of matter
$T^f_{++}={m\over{\lambda x_0^+}}\delta\left(x^+-x^+_0\right)$ in
Kruskal coordinates
\bea
e^{-2\phi}+{Na\over12}\phi
&=&-\lambda^2x^+x^--{N\over48}\log
\left(-\lambda^2x^+x^-\right)\nonumber\\
&&-{m\over\lambda x^+_0}\left(x^+-x^+_0
\right)\theta\left(x^+-x^+_0\right)+C
\>,\label{xliii}
\eea
where $\theta\left(x\right)$ is the step function.
For $x^+>x^+_0$ we have an evaporating black hole solution with a curvature
singularity lying on the critical line $\left(\Omega^{\prime}=0\right)$
\be
\alpha=-\lambda^2x^+\left(x^-+\Delta\right)-{N\over48}\log\left(
-\lambda^2x^+x^-\right)+{m\over\lambda}+C\>,\label{xliv}
\ee
where $\Delta={m\over\lambda^3x_0^+}$ and $\alpha={Na\over24}\left(1-\log
{Na\over24}\right)$. The critical line initially lies behind an apparent
horizon $\partial_+e^{-2\phi}=0$, which is given by the curve
\be
-\lambda^2x^+\left(x^-+\Delta\right)={N\over48}\>.\label{xlv}
\ee
Due to Hawking radiation the apparent horizon recedes and meets the
curvature singularity in a finite proper time. The intersecting point
is located at
\be
x^-_{int}={-\Delta\over 1-{N\over48}\exp\left[-{48\over N}
\left({m\over\lambda}+C-\alpha\right)+1\right]}\>,\label{xlvi}
\ee
\be
x^+_{int}={1\over\lambda^2\Delta}\left\{\exp\left[{48\over N}
\left({m\over\lambda}+C-\alpha\right)+1\right]-{N\over48}\right\}
\>,\label{xlvii}
\ee
after which the singularity becomes naked. Following the reasoning
of \cite{RST,BPP} we shall try to construct a stable solution
which continuously matches the evaporating black hole solution on the null
curve $x^-=x^-_{int}$. A static solution in the asymptotically flat coordinates
$\lambda\hat\sigma^+=\log\lambda x^+,\lambda\hat\sigma^-=-\log\left(-
\lambda\left(x^-+\Delta\right)\right)$ is given by
\be
e^{-2\rho}+{Na\over12}\rho=-\lambda^2x^+\left(x^-+\Delta\right)
-{N\over48}\log\left(-\lambda^2x^+\left(x^-+\Delta\right)\right)+\hat C
\>.\label{xlviii}
\ee
The solution (\ref{xliii}) can be matched to the solution (\ref{xlviii})
if the constant $\hat C$ is
\be
\hat C=-{N\over48}\left(1-\log{N\over48}\right)+\alpha\>.\label{xlix}
\ee
The constraint equations imply the existence
of a shock wave originated at the end-point
\be
T^f_{--}\left(\hat\sigma^-\right)={N\lambda\Delta\over48x^-_{int}}
\delta\left(\hat\sigma^--\hat\sigma^-_{int}\right)\>,\label{l}
\ee
which account for the energy conservation.
 The energy radiated by the black hole is
\bea
&&E_{rad}=\int^{\hat\sigma^-_{int}}_{-\infty}<T^f_{--}\left(\hat\sigma^-
\right)>d\hat\sigma^-=\nonumber\\&&m+\lambda C-\lambda\alpha
-{N\lambda\over48}\left(\log{N\over48}-1\right)-{N\lambda\Delta\over48
x^-_{int}}\>,\label{li}
\eea
and the ADM mass of any static solution (\ref{xlviii}) can be obtained from
the expression \cite{Bilal-Kogan}
\be
M_{ADM}=2e^{2\sigma}\left(\partial_{\sigma}+1\right)\left(\delta\phi
-\delta\phi^2\right)\mid_{\sigma=\infty}\>.\label{ADM}
\ee
One find that
\be
M_{ADM}=\lambda\left(C-C_0\right)\>,\label{lii}
\ee
where $C_0$ corresponds to an arbitrary reference solution. The unradiated
mass before the end-point
\be
{N\lambda\over48}\left(\log{N\over48}-1\right)-
\lambda C_0+\lambda\alpha+
{N\lambda\Delta\over48x^-_{int}}\>,\label{liii}
\ee
coincides with the ADM mass of the remnant $\left(\lambda\left(\hat C
-C_0\right)\right)$ plus the energy of the "thunderpop" (\ref{l}).
Like the BPP model the natural solution is $C_0=\hat C$, for which the
remnant solution turns out to be the ground state. However, unlike the BPP
model, the end-state geometry is not a semiinfinite throat.
The left boundary of the end-state geometry is achieved at the critical
points, for which $\Omega^{\prime}=0\left(i.e,\sigma_{cr}={1\over2}
\log {N\over48}\right)$.
In the proximities of the boundary $\sigma=\epsilon+\sigma_{cr}$ the
metric behaves as
\be
ds^2\sim{e^{2\epsilon}{N\over48}\over{Na\over24}+{Na\over12}\sqrt{a\over2}
\epsilon+\Theta\left(\epsilon^2\right)}\left(-dt^2+d\sigma^2\right)
\>,\label{liv}
\ee
It is not difficult to show that the end-state geometry is still regular
at the boundary although the space-time is no longer geodesically complete.
\section{Towards a solvable model for a radiating 4D black
hole}
The aim of this section is to extend our previous analysis of the CGHS
model to a more general class of 2D dilaton gravity theories.
In particular we want to generalize the field redefinitions
(\ref{xxiv}),(\ref{xxv}) to generate a semiclassical (solvable) model
describing an evaporating spherically symmetric black hole.
The clue for doing this is to realize that the field redefinitions
(\ref{xxiv}),(\ref{xxv}) of the CGHS model can be rewritten in the
form
\be
\rho-\phi=\hat\rho-\hat\phi\>,\label{lv}
\ee
\bea
&&-\int^{y^+}e^{2\left(\rho-\phi\right)}\int e^{-2\left(\rho-\phi\right)}
 G_{++}-\int^{y^-}e^{2\left(\rho-\phi\right)}\int e^{-2\left(\rho-\phi\right)}
 G_{--}=\nonumber\\
 &&-\int^{y^+}e^{2\left(\hat\rho-\hat\phi\right)}\int e^{-2\left(
 \hat\rho-\hat\phi\right)}\hat G_{++}-\int^{y^-}e^{2\left(
 \hat\rho-\hat\phi\right)}\int e^{-2\left(\hat\rho-\hat\phi\right)}
 \hat G_{--}\nonumber\\
 &&+{N\over12}F\left(\hat\phi\right)+{N\over12}M\left(\hat\rho-
  \hat\phi\right)\>,\label{lvi}
  \eea
where $G_{\pm\pm}$ are the null components of the Einstein tensor of the CGHS
model
\be
G_{\pm\pm}=e^{-2\phi}\left(2\partial
^2_{\pm}\phi-4\partial_{\pm}\rho\partial_{\pm}\phi
\right)\>,\label{lvii}
\ee
and $\hat G_{\pm\pm}=G_{\pm\pm}\left(\hat\rho,\hat\phi\right)$.
Moreover the basic properties of the field redefinitions are maintained
 if we generalize the above expressions, for an arbitrary model, in the
 following way
 \be
 j=\hat j\>,\label{lviii}
 \ee
 \bea
&&-\int^{y^+}e^{2j}\int e^{-2j}G_{++}-\int^{y^-}e^{2j}\int e^{-2j}G_{--}
=\nonumber\\
&&-\int^{y^+}e^{2\hat j}\int e^{-2\hat j}\hat G_{++}-\int^{y^-}e^{2\hat j}
\int e^{-2\hat j}\hat G_{--}\nonumber\\
&&+{N\over12}F\left(\hat\phi\right)
+{N\over12}M\left(\hat j\right)\>,\label{lix}
\eea
where $j$
 is a conserved current, $\partial_+\partial_-j=0$, which transforms
as the conformal factor of the metric
\be
j=\rho\ +\ scalar\ terms\>,\label{lxi}
\ee
\be
M\left(\hat j\right)=\hat j+\int^{y^+}e^{2\hat j}\int e^{-2\hat j}
\left(\partial_{y^+}\hat j\right)^2+\int^{y^-}e^{2\hat j}\int e^{-2\hat j}
\left(\partial_{y^-}\hat j\right)^2\>,\label{lx}
\ee
and
\begin{displaymath}
G_{\pm\pm}=T^f_{\pm\pm}
\end{displaymath}
are the classical constrained equations of the model.

For spherically symmetric gravity (1.8) the constrained equations with the
Weyl-effective action are
\be
e^{-2\phi}\left[2\partial^2_{\pm}\phi-4\partial_{\pm}\rho\partial_{\pm}\phi-
2\left(\partial_{\pm}\phi\right)^2\right]=T^f_{\pm\pm}
+T^W_{\pm\pm}\>,\label{lxii}
\ee
and the remaining equations read as
\be
\partial_+\phi\partial_-\phi-\partial_+\partial_-\phi+\partial_+\partial_-
\rho=0\>,\label{lxiii}
\ee
\be
8\partial_+\phi\partial_-\phi-4\partial_+\partial_-\phi+e^{2\left(\rho+\phi
\right)}=0\>.\label{lxiv}
\ee
 From the above equations it is easy to see that the analogue of the CGHS
current $j=\rho-\phi$ is now
\be
j=\rho-{1\over2}\phi-{\lambda^2\over8}\int^{y^+}\int^{y^-}e^{2
\left(\rho+\phi\right)}\>.\label{lxv}
\ee
Due to the non-local term this current should be finally defined by fixing its
value on the Minkowski background $\left(ds^2=-d\sigma^+d\sigma^-,
{e^{-2\phi}\over\lambda}={\sigma^+-\sigma^-\over2}\right)$.
 Following the analogy with the CGHS model
 the natural choice is then
 \be
 j\mid_{Minkowski}={\lambda\over2}\left(\sigma^+-\sigma^-\right)\>.\label{lxvi}
 \ee
 Let us now discuss the possible choices for the function F.
 The model which preserves the classical vacuum, i.e, the analogue of the RST
  model for spherically symmetric gravity, is given by the condition
  \be
  F\left(\phi\right)+M\left(j\right)\mid_{Minkowski}=0\>.\label{lxvii}
  \ee
  A straightforward computation implies that
  \be
  F\left(\phi\right)=-{1\over2}e^{-\phi}\>.\label{lxviii}
  \ee
  The simplest choice is however
  \be
  F\left(\phi\right)=0\>,\label{lxix}
  \ee
  which define the analogue of the BPP model for spherically
  symmetric gravity. Defining the "Kruskal" gauge $\left\{x^{\pm}\right\}$
  by the condition $j=0$ we also have the relation
   $\lambda x^{\pm}=\pm e^{\pm\lambda
  \sigma^{\pm}}$ ) In this gauge the
   equations of motion of the covariant model coincides
  with those of the non-covariant theory:
  \be
  \partial_+\hat\phi\partial_-\hat\phi-\partial_+\partial_-\hat\phi
  +\partial_+\partial_-\hat\rho=0\>,\label{lxx}
  \ee
  \be
  8\partial_+\hat\phi\partial_-\hat\phi-4\partial_+\partial_-\hat\phi+
  \lambda^2e^{2\left(\hat\rho+\hat\phi\right)}=0\>,\label{lxxi}
  \ee
  \be
  e^{-2\hat\phi}\left[2\partial^2_{\pm}\phi-
  4\partial_{\pm}\hat\rho\partial_{\pm}\hat\phi-2\left(
  \partial_{\pm}\hat\phi\right)^2\right]
  =T^f_{\pm\pm}-{N\over48}{1\over\left(x^{\pm}\right)^2}\>,\label{lxxii}
\ee
and therefore the study of the semiclassical equations can be reduced
to that of the classical ones with an effective stress tensor of the form
$T^f_{\pm\pm}-{N\over48}{1\over\left(x^{\pm}\right)^2}$.
We have to remark that, in the absence of matter $T^f_{\pm\pm}=0$,
the above equations admit static solutions in the
"Minkowskian" coordinates $\sigma^{\pm}$. This consistence condition
is a direct consequence of the choice (5.12) in the definition of the
non-local conserved current (5.11).
\section{Conclusions}
In this work we have elaborated a new scheme to generate solvable
models of the CGHS theory in the semiclassical approximation.
The solvability of the covariant one-loop corrected models (3.14)
comes from the fact that all of them can be converted, via suitable
field redefinitions, into a unique non-covariant one-loop theory
(3.1) which inherits the solvability of the classical theory.
The structure of the field redefinitions admit a natural
and consistent
generalization to more general dilaton-gravity models.
We have considered the relevant case of spherically symmetric
gravity and we have defined the analogue of the RST and BPP models.
In the later case the semiclassical equations reduce, in a particular
coordinate system,
to the classical ones with an effective, $\rho$-independent stress tensor.

\section*{Acknowledgements}
We would like to thank M. Navarro and C. F. Talavera for useful
discussions.

 \end{document}